# HandsInAir: A Wearable System for Remote Collaboration

Weidong Huang, Leila Alem and Jalal Albasri
CSIRO ICT Centre
PO Box 76
Epping NSW 1710 Australia
{Tony.Huang, Leila.Alem, Jalal.Albasri}@csiro.au

*Abstract*—We present HandsInAir, a real-time collaborative wearable system for remote collaboration. The system is developed to support real world scenarios in which a remote mobile helper guides a local mobile worker performing a physical task. HandsInAir implements a novel approach to support mobility of remote collaborators. This approach allows the helper to perform gestures without having to touch tangible objects, requiring little environment support. The system consists of two parts: the helper part and the worker part. The two parts are connected via a wireless network and the collaboration partners communicate with each other via audio and visual links. In this paper, we review related work, describe technical implementation of the system and envision future work for further improvements.

## I. INTRODUCTION

Globalization is an inevitable trend in the modern society and collaboration between individuals across the globe and organizations has become an essential part of our daily work. The past decades have seen a fast growing interest among researchers and engineers in developing systems to support remote collaboration [4]. However, most of these systems aim to support collaborations in which individuals play similar roles, such as students working together to finish their group project. Relatively less attention has been given to collaborative activities in which collaborative partners have distinct roles.

In particular, as technologies become increasingly complex, our dependence on external expertise to understand and use the technology is growing rapidly [8]. There are a range of real world situations in which assistance from a remote helper is needed for a local novice worker to accomplish collaborative physical tasks [2]. For example: an ultrasound examination is one of medical checks that require specific expertise to conduct. However, such expertise is often limited in supply and not always available locally. In some cases, there is a need that a remote radiologist guides a non-specialist doctor or nurse manipulating an ultrasound machine to conduct a quality diagnostic ultrasound scan. Other examples include a remote expert providing technical support for a local technician to maintain a piece of equipment; a remote instructor helping a disabled student at home to complete art and craft homework.

It has been widely agreed that one of the main issues with remote collaboration is that there is no longer common ground for them to communicate in a way in which they do when they are co-located [13]. A series of studies have been conducted demonstrating that providing remote collaborators with access to a shared visual space helps to achieve common ground and can be beneficial to the completion of collaborative tasks (e.g., [5, 11]). According to Tang et al. [17], "*a shared visual workspace is one where participants can create, see, share and manipulate artefacts within a bounded space*". For remote collaboration, shared visual spaces are often provided in the form of video views of the workspace of the worker [15].

Further, prior research has indicated that the reason why face-to-face communication is more efficient than video-mediated communication is mainly because in the face-to-face condition, participants are able to perform gestures over the task objects and the gestures are visually available to all participants [11, 12]. This suggests that it is important to support gesturing in the shared visual space for effective remote collaboration.

A number of systems have been developed to support remote guidance providing a share visual space and using the space for gesturing, such as the DOVE system of Ou et al. [15], GestureMan of Kuzuoka et al. [14] and HandsOnVideo of Alem et al. [1]. However, the existing systems generally assume that the helper is confined within a fixed desktop setting and require complex technical and environment support for the helper to perform gestures and for the system to convey the gestures to the worker. How to better support hand gestures when collaborators are fully mobile in non-traditional desktop environments has not been fully explored in the literate.

In this paper, we present HandsInAir, a wearable system for mobile remote guidance. This system implements a novel approach to support the mobility of remote collaborators. This approach allows the helper to perform gestures without having to touch tangible objects. Therefore the system requires little environment support and is ideal when collaborators are mobile.

In the remainder the paper, we first present a theoretic background for our research, with a focus on shared visual space and the role of remote gestures for collaborative physical tasks. Then we briefly review approaches of supporting hand gestures that have been used in previous research, followed by a presentation of our approach. Next we introduce our HandsInAir system with detailed user interface and system specifications. Finally we conclude the paper with a short summary and future work.

## II. THEORETIC BACKGROUND

In performing collaborative physical tasks, people interact with each other via various communication channels. The interpersonal communication can be more effective when collaborators shared greater amount of common ground, which includes mutual knowledge, beliefs, attitudes, expectations, etc.

### A. Shared Visual Space

Previous research has demonstrated the value of shared visual space in achieving common ground [12]. In particular, according to Kraut et al. [13], shared views of a workspace play at least three interrelated roles:
- Maintain situational awareness;
- Aid conversational grounding;
- Promote sense of co-presence.

First, to have a successful collaboration, collaborators need to have on-going awareness of the task and their partner. This awareness can be used to plan what to say and what to do next, serving as a mechanism to coordinate between their verbal utterances and physical actions. Such awareness can be obtained through shared visual views of workspace because collaborators can see what is happening. Second, effective communication largely depends on how much mutual knowledge they have about the task and their partner. More specifically, as stated by Gergle et al. [7], "*speakers form utterances based on an expectation of what a listener is likely to know and then monitor whether the utterance was understood. In return, listeners have a responsibility to demonstrate their level of understanding*". Information needed for building such mutual knowledge can be obtained from shared views of workspace. Third, when collaboration takes place among individuals who are physically distributed, it is important to help collaborators to feel being connected. Enhancing sense of co-presence has proved to be beneficial to the success of collaboration [18]. Shared views of workspace help to promote such sense of "being together" [2].

### B. Remote Gestures

Although a shared visual space is helpful for grounding, or establishing common ground between collaboration partners, it is not feasible, if not impossible, to provide all visual information available to co-located collaborators to remote collaborators, due to bandwidth limitations and limited cognitive capacity of humans [5, 13]. Therefore in developing systems for remote collaboration, it is important to determine what visual information is the most important and make sure this information is provided in an appropriate way.

It has long been acknowledged that the reason why face-to-face communication is generally more effective than video-mediated communication is mainly because in the face-to-face condition, participants are able to perform gestures at the task objects and those gestures are visually available to all participants [10, 11]. Fussell and her colleagues conducted a series of studies on collaborative physical studies and found that not only speech, but also gestures and actions were used for grounding and that the use of gestures improved task performance (e.g., [5, 13, 15]). With access to the shared visual space, helpers allocate most of their attentions on workers' hands and task objects [7, 10]. All these findings indicate that it is important to support remote gestures for developing tools for remote collaboration.

Fussell et al. [6] conducted two studies. The studies were conducted to investigate the role of two types of gestures including pointing and representational gestures and how theses gestures can be effectively conveyed to the remote site. The first study used a system that was mouse-based and supported remote pointing only, while the second study used a system that was pen-based and supported remote drawing of both pointing and representational gestures. The results indicated that only a simple cursor pointing was not enough for effective collaboration, while pen-based drawings of remote gestures resulted in communication and performance being as good as that in co-located collaboration.

Kirk et al. [9] conducted an ethnographic analysis of the nature and role of gestural actions based on a mixed ecology system. This system projected unmediated hand gestures to the remote site to promote collaborative awareness. The analysis led to revealing of a corpus of so-called "gestural phrases". These gesture patterns were used for collaborative physical tasks and they were:
- Flashing Hand
- Wavering Hands
- Mimicking Hands
- Inhabited Hand
- Negating Hand
- Parked Hands

## III. SUPPORTING REMOTE GESTURES

A number of systems have been proposed or developed in the literature, providing a shared visual space and supporting remote gestures using various technologies. In this section, we briefly review approaches that have been used in prior research and present our HandsInAir approach.

### A. Prior approaches

Kurata et al. [16] developed a WACL system in which the worker wears a steerable camera/laser head and the helper can independently set his own viewpoint and point to real objects in the task space with the laser spot. Kuzuoka et al. [14] developed GestureMan systems in which the helper uses joystick to control a mobile robot located on the worker site. Remote gestures are conveyed by the mobile robot through the use of a laser pointer.

Ou et al. [15] developed a DOVE (Drawing over Video Environment) system that integrates gestures of helper into the live video of the worker's workspace. The main feature of the system is that the system allows a remote helper to perform gestures by drawing on video streams of the work environment while providing verbal instructions. In their system, sketching is used to represent gestures and the gestures that are supported include both pointing and representational gestures.

Kirk and Fraser [11, 12] presented a mixed ecology system. This system supports remote gestures using unmediated representations of the helper's hands. In this system, the helper's hands are captured by a video camera and the gestures he made are directly projected onto the desk of the worker to promote mutual awareness between participants.

Most of the early systems for remote collaboration either assume that the workspace of the worker is confined in a fixed desktop setting or support only limited gestures such as pointing. Recently, Alem et al. [1, 3] developed a HandsOnVideo system, aiming to support real world scenarios in which an expert guiding a mobile worker located in a non-traditional-desktop environment. In their system, the helper's hands are captured by a camera and integrated with the scene videos of the workspace. The combination of the hands and workspace scenes are then displayed on a near-eye display worn by the worker and located just above his eyes.

### B. Our Approach

In previous approaches, the helper is often confined within a fixed desktop setting and needs to use, touch or control physical objects to be able to perform gestures that can be conveyed to the worker site. This often requires substantial technical implementation and environment support.

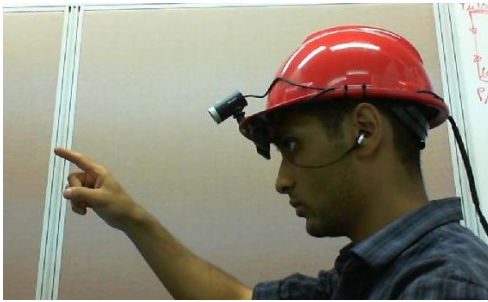

Figure 1. Helper performs gestures in the air.

However, our approach is different. As shown in Figure 1, the helper only needs to wear a helmet mounted with a camera and a near-eye display. Requirements for technical implementation are also minimum compared to other remote collaboration systems. The helper virtually performs gestures in the air with little environment constraint, which is ideal when the helper is mobile.

## IV. SYSTEM OVERVIEW

Our HandsInAir system includes two parts: the helper node and the worker node. The two nodes are connected through a wireless network.

### A. Hardware and Software Implementation

As shown in Figure 2, the hardware used to implement each wearable communication node consists of a helmet mounted with Microsoft Lifecam Webcam on top of, and Vuzix 920 wrap near-eye display beneath the brim. Both the worker and the helper interfaces use the same set of hardware. The camera is used to capture the workspace scenes or the helper's hands. The near-eye display is used to display the scene videos for the helper and visual instructions for the worker. Both the peripheral devices are connected to a laptop worn by the user. There is an audio connection between the two sides to support verbal communications.

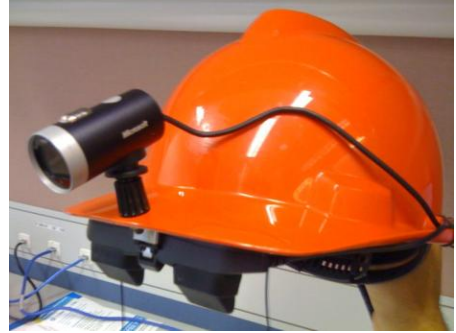

Figure 2. User interface.

The system's software implementation is developed in C++ on Windows XP machines utilizing a number of open source libraries to perform the various networking and computer vision functions as required.

Both the worker and helper nodes simultaneously act as a video server and a video client. The worker node acts as a server sending local camera feeds of the workspace and as a client to the helper node receiving video feeds of the helper's hands. Likewise the helper station acts as a video client receiving workspace feeds and sending video feeds containing the helper's hands.

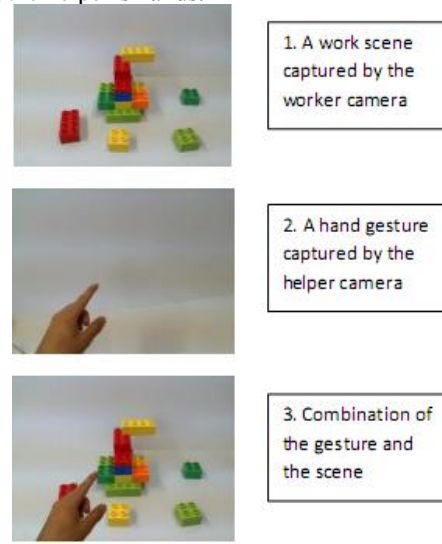

Figure 3. Illustration of combining a hand gesture and a workspace scene.

The Intel OpenCV open source computer vision library is used to implement an Adaptive Skin Detector, which extracts the helper's hands from video feeds of the helper camera and combines them with corresponding video feeds of the workspace (see Figure 3). This detector is also used to display the combined videos on the near-eye displays of the helper and the worker.

Network connections are realized low level by opening up streaming connections as both machines simultaneously send and receive a sequence of images. The images are compressed with JPEG compression prior to sending and decompressed upon receipt using the open source IJG (Independent JPEG Group) LibJPEG library to avoid sending costly raw image data and to maintain a real time frame rate at both nodes.

*B.  How the System Works*

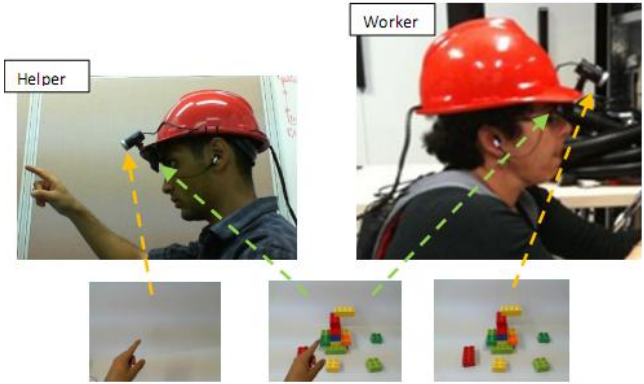

Figure 4.  Camera captures and the content of near-eye displays.

How the system works is illustrated in Figure 4. Once a connection is established the system initializes two video streams between the nodes. First the scene video from the worker camera is fed to the helper node and displayed on the near-eye display. The helper performs gestures which are captured the helper camera. The hands are extracted without the background and combined with the scene video. What is shown on the helper's near-eye display is continuously updated with the combination. In other words, the helper is able to see his hands performing gestures at the task artefacts on the display. The combination of hands and the scene is also sent to the worker side and displayed on the near-eye display of the worker.

## V.  CONCLUSION AND FUTURE WORK

In this paper, we have presented HandsInAir, a new real-time wearable system for remote collaboration. The system employs a novel approach that supports mobility of remote collaborators and remote gestures. This approach enables the helper to perform hand gestures in the air without having to touch any tangible objects. The system is lightweight, easy to set up, intuitive to use, and requires little environment support. We believe that HandsInAir has great potential to be used in a wide range of real world application domains, such as telemedicine and remote repair.

We are planning a formal usability evaluation to gain in-depth understanding of the usability and usefulness of HandsInAir. In the future, we will investigate possibilities of reorganizing the camera and the near-eye display in the user interfaces to see whether the whole system could be more comfortable to wear and more user-friendly to use.